\documentstyle[12pt]{article}

\def\be{\begin{equation}}
\def\ee{\end{equation}}
\def\bea{\begin{eqnarray}}
\def\eea{\end{eqnarray}}

\newcommand{\bean}{\begin{eqnarray*}}
\newcommand{\eean}{\end{eqnarray*}}
\newcommand{\eqn}[1]{(\ref{#1})}

\def\cstar{C$^*$-algebra }

\def\l{\lambda}

\def\ca{{\cal A}}
\def\cb{{\cal B}}
\def\cc{{\cal C}}

\def\ch{{\cal H}}

\def\ck{{\cal K}}

\def\cp{{\cal P}}

\def\br{\begin{thebibliography}{abc}}
\def\er{\end{thebibliography}}
\def\rf{\bibitem}

\def\cstar{C$^*$-algebra }

\newcommand{\I}{\mbox{\rm I} \hspace{-0.5em} \mbox{\rm I}\,}

 \newcommand{\complex}{
        \mbox{C \hspace{-1.16em} \raisebox{-0.018em}{\sf l}}\;}
\def\iff{\Leftrightarrow}

\def\bar#1{\overline{#1}}

\def\Hat#1{\rlap{\kern.10em$\widehat{\phantom G}$}#1}
\def\HAt#1{\rlap{\kern.05em$\widehat{\phantom G}$}#1}

\def\czp#1{\rlap{\kern.1em$\widehat{\phantom{G\vrule height.8em}}$}#1{}}
\def\Czp#1{\rlap{\kern.05em$\widehat{\phantom{G\vrule height.8em}}$}#1{}}

\newcommand{\sect}[1]{\setcounter{equation}{0}\section{#1}}
\newcommand{\subsect}[1]{\subsection{#1}}

\newcommand{\nn}{{^n}}
\newcommand{\inff}{{^\infty}}

\def\ca{{\cal A}}
\def\ch{{\cal H}}
\def\pr{{\cal I}}

\renewcommand{\thefootnote}{\fnsymbol{footnote}}
\def\fn{\footnote}
\def\sxn#1{\bigskip\medskip \sect{#1} \smallskip
                                                 }
\def\subsxn#1{\medskip \subsect{#1} \smallskip
                                                }

\begin{document}

\thispagestyle{empty}
\setcounter{page}{0}

\begin{flushright}

ESI 240 (1995)\\
DSF-T-36/95\\
UICHEP-TH/95-7\\
hep-th/9507148\\

\hfill July 1995
\end{flushright}

\vspace{.25cm}

\centerline {\Large NONCOMMUTATIVE LATTICES}
\vspace{.5cm}
\centerline{ \Large AND THEIR CONTINUUM LIMITS}
\vspace{1cm}
\centerline {\large G. Bimonte$^{1,}$\fn{Address after September
$1^{\rm st}: $ Departamento de Fisica Teorica, Facultade de Ciencias,
Universitad de Zaragoza, 50009 Zaragoza, Spain.},
                    E. Ercolessi$^2 $,
                    G. Landi$^{3,4}$,}
\vspace{2mm}
\centerline{\large  F. Lizzi$^{4,5}$,
                    G. Sparano$^{4,5}$,
                    P. Teotonio-Sobrinho$^6$}
\vspace{1.5cm}
\centerline {\it The E. Schr\"odinger International Institute for
Mathematical Physics,}
\centerline{\it Pasteurgasse 6/7, A-1090 Wien, Austria.}
\vspace{2mm}
\centerline {\it $^1$ International Centre for Theoretical Physics,
P.O. Box 586, I-34100, Trieste, Italy.}
\vspace{2mm}
\centerline {\it $^2$ Dipartimento di Fisica and INFM, Universit\`a di
Bologna,}
\centerline{\it Via Irnerio 46, I-40126, Bologna, Italy.}
\vspace{2mm}
\centerline{\it $^3$ Dipartimento di Scienze Matematiche,
Universit\`a di Trieste,}
\centerline{\it P.le Europa 1, I-34127, Trieste, Italy.}
\vspace{2mm}
\centerline {\it $^4$ INFN, Sezione di Napoli, Napoli, Italy.}
\vspace{2mm}
\centerline {\it $^5$ Dipartimento di Scienze Fisiche, Universit\`a di
Napoli,}
\centerline{\it Mostra d' Oltremare, Pad. 19, I-80125, Napoli, Italy.}
\vspace{2mm}
\centerline {\it $^6$ Dept.\ of Physics, Univ.\ of  Illinois at Chicago,}
\centerline{\it 60607-7059, Chicago, IL, USA.}
\vspace{.2cm}
\begin{abstract}
We consider finite approximations of a topological space $M$ by
noncommutative lattices of points. These lattices are structure
spaces of noncommutative
$C^*$-algebras
which in turn approximate the
algebra $\cc(M)$ of continuous functions on $M$.
We show how to recover the space $M$ and the algebra $\cc(M)$ from a
projective system of noncommutative lattices and an inductive system
of noncommutative $C^*$-algebras, respectively.
\end{abstract}

\newpage
\setcounter{page}{1}

\renewcommand{\thefootnote}{\arabic{footnote}}
\setcounter{footnote}{0}

\sxn{Introduction}\label{se:1}

Physical models are usually given by a suitable carrier
(configuration or phase) space together with a dynamics on it.
Nontrivial topological properties of the
carrier space may have deep consequences independently of any
particular dynamics. Topological solitons in
semiclassical  physics and the existence of inequivalent
quantizations ($\theta $-states) are two familiar examples.

Realistic physical models, however, are often too complicated to be
solved exactly. One is obliged to use approximation methods. In this
context, lattice theories have become a standard tool in the study of
non perturbative aspects of physical models, especially of gauge
theories. An important development would consist in trying to
accommodate in the lattice picture also the topological aspects of
the models. However, usual lattice theories  \cite{lattice}
generally mix aspects of the carrier space and of the dynamics.
Moreover, it is not obvious how topological information can be encoded.
One can ask, for example, how the underlying continuum space (-time)
can be recovered from the discrete data.
In typical models, the only topological information refers to
the underlying space (-time) and is that of nearest neighbors as
encoded in the Hamiltonian. This does not define
per se a notion of limit from which the space (-time) itself is
recovered. A more substantial problem is that
this incomplete topological information has no bearings
on the configuration space of fields which is topologically trivial.

In a previous paper \cite{comp} we made an attempt to clarify some of
these questions. In particular,
we have shown that the notions of inductive and
projective limit allow to recover a topological space $M$ from
a sequence of lattice approximations. We found that although the
lattices had a trivial topology, namely were sets of isolated
points, from the limit $Q^\infty$ one could still recover the space being
approximated, and its algebra of continuous functions. However, due to
the too simple topological properties of the model, the construction
could not be complete.
On one side, at a fixed level of approximation,
all topological information on $M$ was lost. On the other side, $Q^\infty$
had a universal character, in the sense that it was the same
for all spaces $M$. In order to recover  the latter it was necessary
the additional input of a projection from $Q^\infty$ onto $M$. This
projection, though provided by the specific system of lattices
approximating $M$, was not definable directly from the knowledge of
$Q^\infty$ . In this paper we will show how these
drawbacks can be overcome with the use of
noncommutative lattices, i.e., $T_0$ topological spaces with a
finite (or countable) number of points
\cite{BBET,NCL} as opposed to
ordinary Hausdorff lattices.

This paper is part of the program initiated in
\cite{BBET,NCL} where we investigate
these lattices, and their geometry from an algebraic point of view.
The starting point is a paper of Sorkin \cite{So}, where an approximation
scheme for topological spaces is proposed. The idea is to approximate the
space
with a finite (or countable) topological space which is however not Hausdorff,
but only $T_0$. Those space are Partially Ordered Sets (posets) and the
topology is
determined by
the partial order. This approximation greatly improves the
traditional (Hausdorff) finite approximation. With a finite number of points
it is possible to reproduce correctly some relevant topological properties
of
the space being approximated. In the same paper Sorkin pointed out that the
notion of projective system gives a well defined notion of continuum limit,
from which the original space can be reconstructed.

In \cite{BBET,NCL} we developed the essential tools for doing quantum
physics on posets. It was observed that posets are genuine
noncommutative spaces (noncommutative lattices).
Indeed, one can associate with any poset $P$ a
noncommutative algebra ${\cal A}$ of `operator valued'
continuous functions, from which $P$ can be
reconstructed as a topological spaces
much in the same way in which the Gel'fand-Naimark
procedure allows to reconstruct a Hausdorff topological space from the
commutative algebra of continuous functions defined on it.
Thus the term noncommutative lattice, which we use interchangeably with
the term poset.
The algebraic framework provides the way to discuss
quantum mechanical and field theoretical models, in the spirit of
Connes' noncommutative geometry \cite{Co}. It is important to notice
that these noncommutative spaces, although have a poorer geometry than
their continuum counterpart (but not a trivial one), present an
extremely rich algebraic structure. In this way, topological
information enter non trivially at all stages of the construction. In
\cite{NCL} we have also explicitly shown how not trivial topological
effects are captured by these topological lattices and their algebras,
by constructing algebraically the $\theta$-quantizations of a particle
on the noncommutative lattice approximation of a circle.

In this paper we analyze the continuum limit of a sequence of
noncommutative lattices and of their related noncommutative algebras.
The picture which emerges is the following. Given a topological space
$M$ with a sequence of finer and finer cellular decompositions, we
construct a sequence of noncommutative lattices by associating
any such a lattice $P^n$ to each decomposition.
The natural projections $\pi^{(n,m)}$ among the lattices in the
sequence define a projective
system. The projective limit $P^\infty$ is a quasi fiber
bundle\footnote{The fibers above two distinct points do not need to be
homeomorphic and not even of the same dimension.} on $M$,
and $M$ itself is homeomorphic to the quotient of
$P^\infty$ by the equivalence relation defined by the projection from
$P^\infty$ to $M$. As we said above, this quotient is naturally defined in
$P^\infty$ which is then `not too different' from $M$ itself.
This
construction is then dualized. To each noncommutative lattice $P^n$ we
associate a noncommutative algebra $\ca_n$ whose structure space,
$\widehat{\ca}_n$, is $P^n$. By pull-back, the projections
$\pi^{(n,m)}$ define immersions $\Phi_{(m,n)}$ among the algebras,
giving rise to a structure of inductive system. The inductive limit
$\ca_\infty$ results to be the dual of $P^\infty$, namely
$\widehat{\ca}_\infty = P^\infty$. The algebra of continuous functions on $M$
is the center of $\ca_\infty$. Finally, we consider, with an analogous
construction, an inductive system of Hilbert spaces $\ch^n$ on which
the algebras $\ca_n$ act. The Hilbert space
$L^2(M)$ is recovered as a suitable subspace
of the inductive limit $\ch^\infty$.

\sxn{ Continuum Limit of Noncommutative Lattices}

In this section we introduce the notion of continuum limit of a
sequence of finite non Hausdorff topological approximations of a
topological space $M$. In \cite{NCL} we used noncommutative lattices
to approximate topological spaces. These lattices were constructed
starting from open coverings through a quotienting procedure. In this
section we will modify this construction starting instead from a
cellular decomposition. This point of view is more suitable for the
analysis of the continuum limit of a sequence of such lattices and
their related algebras.

Consider a topological space $M$ of dimension $d$ which admits a
locally finite cellular decomposition,
$\Sigma = \{ S_{m,\alpha} \ \alpha\in I\subset N, \ 0 \le m \le d\}$. For
convenience we will use cubic cells and $S_{m,\alpha}$ will then be
closed cubes of dimension $m$. The lattice
$P^{\Sigma}(M)$
is constructed by
associating a point $p_{m\alpha}$ to each cube $S_{m\alpha}$. In the
following, if convenient for clarity, we will use interchangeably
$p_{m\alpha}$ and $S_{m\alpha}$ to indicate either a point of the
lattice or a cube. We
introduce now in $P^{\Sigma}(M)$ the partial order relation,
$\preceq$, given by the inclusion of cubes
\be
p_{m\alpha}\preceq p_{n\beta}
\iff S_{m\alpha}\subseteq S_{n\beta}.
\ee
This partial order relation defines naturally a $T_0$ topology on
$P^{\Sigma}(M)$
\footnote{A $T_0$ topology on a space $M$
is a topology such that, given any two points of $M$,
there exists an open set containing one point but not the other.}.
The topology is generated by the open sets ${\cal O}(S_{m\alpha})$
defined as
\be
{\cal O}(S_{m\alpha}) = \{S_{m'\alpha'}: ~~
S_{m'\alpha'}\subseteq S_{m\alpha} \}
\ee
Any such topological space is also called poset (partially ordered
set) and is usually represented  by a Hasse diagram \cite{So}. We
will use a different diagrammatic representation which is more
transparent in some of our examples and which is described in fig. 2.1.

\bigskip

\noindent{\bf Example 2.1.}
As a simple example consider the interval $[0,1]$ and the cubic
decomposition
\be
\Sigma_2 = \{S_{01} = 0, \ S_{02} = {1\over 2}, \ S_{03} = 1, \
           S_{11} = [0,{1\over 2}], S_{12} = [{1\over 2}, 1] \}.
\ee
The corresponding diagram is shown in fig. 2.1.

\begin{figure}[htb]
\begin{center}
\unitlength=1.00mm
\special{em:linewidth 0.4pt}
\linethickness{0.4pt}
\begin{picture}(146.00,36.00)
\put(65.00,30.00){\line(-1,0){60.00}}
\put(65.00,20.00){\makebox(0,0)[cc]{$1$}}
\put(35.00,20.00){\makebox(0,0)[cc]{$\frac{1}{2}$}}
\put(5.00,20.00){\makebox(0,0)[cc]{$0$}}
\put(5.00,36.00){\makebox(0,0)[cc]{$S_{01}$}}
\put(20.00,36.00){\makebox(0,0)[cc]{$S_{11}$}}
\put(35.00,36.00){\makebox(0,0)[cc]{$S_{02}$}}
\put(50.00,36.00){\makebox(0,0)[cc]{$S_{12}$}}
\put(65.00,36.00){\makebox(0,0)[cc]{$S_{03}$}}
\put(85.00,30.00){\circle*{2.00}}
\put(115.00,30.00){\circle*{2.00}}
\put(145.00,30.00){\circle*{2.00}}
\put(89.00,30.00){\line(1,0){6.00}}
\put(103.00,30.00){\line(1,0){8.00}}
\put(119.00,30.00){\line(1,0){6.00}}
\put(133.00,30.00){\line(1,0){8.00}}
\put(85.00,36.00){\makebox(0,0)[cc]{$p_{01}$}}
\put(115.00,36.00){\makebox(0,0)[cc]{$p_{02}$}}
\put(145.00,36.00){\makebox(0,0)[cc]{$p_{03}$}}
\put(99.00,36.00){\makebox(0,0)[cc]{$p_{11}$}}
\put(129.00,36.00){\makebox(0,0)[cc]{$p_{12}$}}
\put(35.00,10.00){\makebox(0,0)[cc]{$(a)$}}
\put(115.00,10.00){\makebox(0,0)[cc]{$(b)$}}
\put(65.00,28.00){\line(0,1){4.00}}
\put(35.00,28.00){\line(0,1){4.00}}
\put(5.00,28.00){\line(0,1){4.00}}
\put(99.00,30.00){\makebox(0,0)[cc]{$\times $}}
\put(129.00,30.00){\makebox(0,0)[cc]{$\times $}}
\end{picture}
\end{center}
{\footnotesize {\bf Fig. 2.1} Figure (a) shows the cell decomposition
$\Sigma_2$ of the
interval. Figure (b) represents the corresponding noncommutative lattice
$P^2$.
Any two points immediately related are joined by a line. The ordering
relation
is given by the convention that points denoted by  $\times $ are preceded
by
points denoted by $\bullet $}.
\end{figure}

The corresponding noncommutative lattice is
$P^2(I) =
(p_{01},p_{02},p_{03},p_{11},p_{12},)$, the partial order reads
\be
p_{01}\preceq p_{11}~,~~ p_{02}\preceq p_{11},~,~~
p_{02}\preceq p_{12},~,~~ p_{03}\preceq p_{12},~,~~ \label{2.5}
\ee
where we have omitted writing the relations $p_{ij}\preceq p_{ij}$.

In an analogous way, one can construct the noncommutative lattice $P^n(I)$
corresponding to the cubic decomposition $\Sigma_n$
in which the interval $I$ is divided in $2^{n-1}$ subintervals of
equal length. In the following,
when there is no ambiguity we shall write simply $P^n$ instead of
$P^n(I)$.

In the language of partially ordered sets, the smallest open set $O_p$
containing a point $p \in P(M)$ consists of all $q$ preceding
$p$: $O_p = \{q \in P(M) \; : \; q \preceq p \}$.
For instance, in fig. 2.1, this rule gives
$\{p_{01}, p_{02}, p_{11}\}$ as the smallest open set containing
$p_{11}$. These open sets generate the topology.

In order to recover $M$ through a limiting procedure, a natural
framework is the  one of projective systems which we now briefly
recall.
A  {\em projective} (or {\em inverse}) {\em system}  of  topological spaces
is a family of topological spaces $Y\nn , n \in {\bf N}$
\footnote{More generally, the index $n$ could be taken in any directed set. }
together with a family of continuous projections
$\pi^{(m,n)}:Y^{m}\to Y\nn,\ n\leq m$, with the
requirements that $\pi^{(n,n)}~=~\I,\ \pi^{(n,m)}=\pi^{(n,p)}\pi^{(p,m)}$.
The projective limit
$Y^\infty$ is defined as the set of coherent sequences, that is the set of
sequences $\{x\nn\in Y\nn\}$ with $x\nn=\pi^{(m,n)}(x^{m})$.
There is a natural projection
$\pi^{n} : Y^\infty \rightarrow Y\nn $ defined as:
\be
\pi^{n}(\{x^{m}\in Y^{m} \})= x^{n}~.
\ee
The space $Y^{\infty}$ is given a topology, by declaring that a set
${\cal O}^{\infty}\subset Y^{\infty}$ is open iff it is the inverse image of
an open set belonging to some $Y^{n}$ or a union (finite or
infinite) of such sets.

Let us consider then a sequence $\Sigma_n= \{S^n_{m\alpha}, \ \alpha\in I^n\}$
of finer and finer cubic decompositions of $M$,
obtained in the following way: $\Sigma_{n+1}$ is obtained from
$\Sigma_n$ by an even subdivision of its cubes, the precise meaning
of ``even" is that for any point $x\in M$
and any open set ${\cal O}_x$ containing $x$, there must exist a
level of approximation such that all cubes containing $x$ will be
contained in ${\cal O}_x$ from that level on
\fn{We are really using decompositions which are `fat' in the sense
of \cite{CMS}.}
\be
\forall~ x~ {\rm and}~ \ \forall~ {\cal O}_x \ni x,\ \exists~ m~
{\rm ~such~ that~}~ \forall~ n\geq m~,
S\nn_\alpha\ni x\  \Rightarrow
\ S\nn_\alpha\subset{\cal O}_x~. \label{reqseq}
\ee
The structure of projective system is given by the projections
\be
\pi^{(m,n)} : P^m \rightarrow P^n \ \ m>n
\ee
which associate to a cube of $\Sigma_m$ the corresponding cube of
$\Sigma_n$ from which it comes, namely the lowest dimensional cube of
$\Sigma_n$ to which it belongs. Fig. 2.2 shows the case of the interval.

\begin{figure}[htb]
\begin{center}
\unitlength=1.00mm
\special{em:linewidth 0.4pt}
\linethickness{0.4pt}
\begin{picture}(141.00,54.33)
\put(65.00,40.00){\line(-1,0){60.00}}
\put(80.00,40.00){\circle*{2.00}}
\put(110.00,40.00){\circle*{2.00}}
\put(140.00,40.00){\circle*{2.00}}
\put(84.00,40.00){\line(1,0){6.00}}
\put(98.00,40.00){\line(1,0){8.00}}
\put(114.00,40.00){\line(1,0){6.00}}
\put(128.00,40.00){\line(1,0){8.00}}
\put(65.00,20.00){\line(-1,0){60.00}}
\put(80.00,20.00){\circle*{2.00}}
\put(140.00,20.00){\circle*{2.00}}
\put(106.00,20.00){\line(-1,0){22.00}}
\put(114.00,20.00){\line(1,0){22.00}}
\put(140.00,36.00){\vector(0,-1){12.00}}
\put(80.00,36.00){\vector(0,-1){12.00}}
\put(110.00,36.00){\vector(0,-1){12.00}}
\put(124.00,36.00){\vector(-3,-4){9.00}}
\put(94.00,36.00){\vector(1,-1){12.00}}
\put(30.00,10.00){\makebox(0,0)[cc]{$(a)$}}
\put(110.00,10.00){\makebox(0,0)[cc]{$(b)$}}
\put(5.00,36.00){\vector(0,-1){12.00}}
\put(65.00,36.00){\vector(0,-1){12.00}}
\put(50.00,36.00){\vector(0,-1){12.00}}
\put(20.00,36.00){\vector(0,-1){12.00}}
\put(35.00,36.00){\vector(0,-1){12.00}}
\put(5.00,18.00){\line(0,1){4.00}}
\put(65.00,18.00){\line(0,1){4.00}}
\put(65.00,38.00){\line(0,1){4.00}}
\put(35.00,38.00){\line(0,1){4.00}}
\put(5.00,38.00){\line(0,1){4.00}}
\put(94.00,40.00){\makebox(0,0)[cc]{$\times $}}
\put(124.00,40.00){\makebox(0,0)[cc]{$\times $}}
\put(110.00,20.00){\makebox(0,0)[cc]{$\times $}}
\put(35.00,48.00){\circle*{0.67}}
\put(35.00,51.00){\circle*{0.67}}
\put(35.00,54.00){\circle*{0.67}}
\put(110.00,48.00){\circle*{0.67}}
\put(110.00,51.00){\circle*{0.67}}
\put(110.00,54.00){\circle*{0.67}}
\end{picture}
\end{center}
{\footnotesize {\bf Fig. 2.2.} The subdivision of a cell is illustrated
in (a). The respective projection for the noncommutative lattices are
indicated
by arrows in (b) }
\end{figure}

We call $P\inff(M)$ the projective limit of this projective system.
A point in $P^\infty(M)$ is nothing
but a coherent sequence $\{p\nn\}$ of cubes, namely a sequence such
that $p^{n+1}\subseteq p\nn$.
There exists a natural
projection $\pi : P^{\infty }(M) \rightarrow M$. It is defined as
follows:
\be
\pi(\{p\nn\}) =   \bigcap_n  p\nn. \label{prob}
\ee
In this manner, we get a unique point of $M$. That this point is
unique is a consequence of condition (\ref{reqseq}). Due to this
projection, $P^\infty$ results to be a quasi fiber bundle on $M$,
namely a fiber bundle such that the fibers can change
from point to point.
A further characterization of $P^\infty$ is given by the  following
observations.
Let us introduce the set $M_0\subset M$ which is the union of all
closed $d - 1$ cubes of all levels of approximations, namely
\be
M_0= \bigcup_n S^n_{d-1,\alpha}.
\ee
Then $P^\infty$ is a quasi fiber bundle on $M$ with the following
properties:
\begin{itemize}
\item[1.] Only the fibers above points of $M_0$ have more than one point.
Suppose in fact that there are two distinct points, $p=\{p^n\}$ and
$p'=\{p'^n\}$ , of $P^\infty(M)$, projecting on the same point $x$ of
$M$. Since $p$ and $p'$ are different points of
$P^\infty$, there
must be a level $n$ such that $p^n\neq p'^n$. However,
since $p$ and $p'$
project on the same point, the intersection of $p^n$ and $p'^n$ must
be not empty.  But then
the intersection can only  belong
to a lower dimensional cube of the
boundary of both $p^n$ and $p'^n$ and thus $x$ belongs to $M_0$.

\item[2.] Every fiber is a connected partially ordered set with a
       minimal point, namely a point $p$ such that $p\preceq q$ for
       all the elements $q$ in that fiber. This will be clear from
       the proof of Proposition 2. later on.

\item[3.] As it can be seen in simple examples, in general
the nontrivial fibers are not all isomorphic.
\end{itemize}

\bigskip

\noindent{\bf Example 2.2.}
In the example of the interval the fibers of $P^\infty$  are
characterized as follows.
$I_0$ is the set of all points of the interval of coordinates
$m/2^n$, with $m$ and $n$ arbitrary non-negative integers.
The fibers above points of $I - I_0$ have only one point.
The fibers over the points of $I_0$ have three points and the
structure of the poset $V$ (except the boundary points of the
interval whose fibers have only two points). The poset $V$ is the
poset shown in fig. 2.3 above the point $x_3$.

\begin{figure}[htb]
\begin{center}
\unitlength=1.00mm
\special{em:linewidth 0.4pt}
\linethickness{0.4pt}
\begin{picture}(101.00,65.00)
\put(40.00,50.00){\circle*{2.00}}
\put(40.00,46.00){\line(0,-1){7.00}}
\put(55.00,50.00){\circle*{2.00}}
\put(70.00,50.00){\circle*{2.00}}
\put(70.00,61.00){\line(0,-1){7.00}}
\put(70.00,46.00){\line(0,-1){7.00}}
\put(85.00,50.00){\circle*{2.00}}
\put(100.00,50.00){\circle*{2.00}}
\put(100.00,61.00){\line(0,-1){7.00}}
\put(40.00,10.00){\makebox(0,0)[cc]{$x_1$}}
\put(55.00,10.00){\makebox(0,0)[cc]{$x_2$}}
\put(70.00,10.00){\makebox(0,0)[cc]{$x_3$}}
\put(85.00,10.00){\makebox(0,0)[cc]{$x_4$}}
\put(100.00,10.00){\makebox(0,0)[cc]{$x_5$}}
\put(11.00,50.00){\makebox(0,0)[cc]{$P^\infty $}}
\put(11.00,20.00){\makebox(0,0)[cc]{$M$}}
\put(11.00,45.00){\vector(0,-1){20.00}}
\put(5.00,35.00){\makebox(0,0)[cc]{$\pi $}}
\put(40.00,20.00){\line(1,0){60.00}}
\put(70.00,65.00){\makebox(0,0)[cc]{$\times $}}
\put(40.00,35.00){\makebox(0,0)[cc]{$\times $}}
\put(100.00,65.00){\makebox(0,0)[cc]{$\times $}}
\put(70.00,35.00){\makebox(0,0)[cc]{$\times $}}
\put(40.00,20.00){\circle{1.33}}
\put(55.00,20.00){\circle{1.33}}
\put(70.00,20.00){\circle{1.33}}
\put(85.00,20.00){\circle{1.33}}
\put(100.00,20.00){\circle{1.33}}
\end{picture}
\end{center}
{\footnotesize {\bf Fig. 2.3.} The figure shows some of the fibers of
$P^\infty $. The fibers above points in $I_0$
can be
identified with the poset V.
For points in the complement of $I_0$, the fibers
consist of only one
point.}
\end{figure}

In \cite{comp}, where we considered Hausdorff lattices, it was found
that the projective limit, $Q^\infty(M)$, was bigger than the starting
space $M$, and $M$ was recovered from $Q^\infty(M)$ as a quotient
defined by a projection from $Q^\infty(M)$ to $M$ similar to the one
in \eqn{prob}. Furthermore, $Q^\infty(M)$ was a universal space,
namely it was a Cantor set independently
on the starting space $M$ and the information about $M$ was contained in the
projection. As in the case of Hausdorff lattices,
$P^{\infty}$ is bigger than $M$
and $M$ is recovered from $P^\inff$ as a quotient \footnote{This is the same
result as the one found in \cite{So}. The difference is that we
consider simplicial decompositions instead of open coverings.}.
However, unlike the Hausdorff case, $P^\infty$ is not universal but is
uniquely associated to $M$. In fact, as we will show in a moment, the
quotient can now be defined also intrinsically from $P^\infty$ without
using the projection. We shall first prove the following.

{\it Proposition 1.} The space $M$ is homeomorphic to
$P^\infty /\sim $,
where $\sim $ is the equivalence relation defined by the projection
$\pi$ in (\ref{prob}).

{\it Proof.}
\begin{itemize}
\item[~] We prove that: $1.$ $\pi$ is continuous; $2.$ the
quotient $P^\infty/\sim$ is actually homeomorphic to $M$.
\item[~] {\bf $1.$ $\pi$ is continuous } \\
We have to  show that the inverse image of an open set $B$ in $M$ is open in
$P^\infty(M)$. Let $p= \{S^n_{m,\alpha}(p)\}$ be a point belonging to
$\pi^{-1}(B)$ and let $x = \pi(p)$. Because of condition (\ref{reqseq}) on
the sequence of cubic  decompositions there exists a $j\in N$ such that $n > j$
implies that all cubes $S^n_{m,\alpha}$ containing $x$ are all contained in
$B$. Consider then
${\cal O}_n^\infty = (\pi^n)^{-1}({\cal O}(S^n_{m,\alpha}(p)))$
with $n>j$
which is an open set of $P^\infty(M)$ containing $p$. ${\cal O}_n^\infty$
is also
entirely contained in $\pi^{-1}(B)$; all its points are coherent sequences
whose representatives at level $n$ are cells fully contained in
$S^n_{m,\alpha}(p)$ and since the cube $S^n_{m,\alpha}(p)$ is fully
contained
in $B$ they project to points in $B$.

\item[~]{\bf $2.\ P^\infty/ \sim $  is homeomorphic to $M$}

To prove that the topology of $M$ is equivalent to the quotient
topology on $P\inff /\sim$, it is then sufficient to show that the
inverse image of a subset  of $M$, which is {\em not}  open, is not
open in $P\inff$ as well. Consider then the set $\pi^{-1}(B)\subset
P\inff$, with $B \subset M$ {\em not} open. We will show that the
assumption that $\pi^{-1}(B)$ is open leads to a contradiction. The
statement that $B$ is not open in the topology of $M$ is equivalent to
saying that there exists a sequence of points $\{ x_i \}$ of $M$, not
belonging to $B$, which converges to a point $x \in B$. From this
sequence we will extract a particular subsequence $\{ y_j \}$, still
converging to $x$. We first introduce a countable basis of open
neighborhoods for $x$, namely a countable family $\{ {\cal O}_i\}$ of
decreasing open sets containing $x$. Let us start with $ {\cal O}_1$.
Due to condition (\ref{reqseq}), there are one or more $d$-cubes
$S_{d,\alpha}^{n(1)} \subset { \cal O}_1$,
with $ S_{d,\alpha}^{n(1)} \ni x$.
At least one of these $d$ cubes, call it
$S_{d,\alpha(1)}^{n(1)}$, will contain an infinite number of elements
of the sequence $\{x_i\}$. Then, choose $y_1$ to be any one of these
elements. At the next level $2$, there will again be at least one
$d$-cube $  S_{d,\alpha(2)}^{n(2)} \subset S_{\alpha(1)}^{n(1)} $,
with $ S_{d,\alpha(2)}^{n(2)}$ still containing an infinite number of
elements of the sequence ${x_i}$. Again choose $y_2$ as any one of
these elements. By iterating this procedure, we obtain the sequence
$\{y_j \}$, which, being extracted from the original sequence, still
converges to $x$. Moreover, $y_j \in S_{d,\alpha(j)}^{n(j)} $ and
$\{S_{d,\alpha(j)}^{n(j)} \}$ is a coherent sequence
\footnote{A little
care must be taken as $n(j)$ may not coincide with the level of the
lattice.} which thus defines a point $q\in Q \inff$. By construction
$\bigcap_j S_{d,\alpha(j)}^{n(j)} = x$, and consequently $\pi(q)=x$.

Since $\pi^{-1}(B)$ is assumed to be open and
recalling how the topology of $Q\inff$ is defined,
there will be a $\bar{j}$ such that
${\cal O}_n^\infty = (\pi^{n(j)})^{-1}({\cal O}(S^{n(j)}_{m,\alpha(j)}(p)))
\subset \pi^{-1}(B)$ for $j \geq {\bar j}$.
But then also
$\pi^{-1}(y_j)$, with $ j \geq \bar{j}$ , must belong to
$\pi^{-1}(B)$ and this implies,
contrary to the hypothesis on the sequence $\{x_i\}$, that $y_j\in B$.
\end{itemize}

We show finally that
the equivalence relation $\sim$ can be defined,
without using any reference to the projection $\pi$, in the
following way.

{\it Proposition 2.} For any $q_1, q_2\in P^\infty,$
\be
q_1\sim q_2 \Leftrightarrow \exists q_3 \in P^\infty :
q_3\preceq q_1, q_3\preceq q_2.
\ee

{\it Proof.}
\begin{itemize}
\item[~] The $\Rightarrow$ part can be proved by the explicit construction
of $q_3$,
for $q_1\sim q_2$. The coherent sequence representing $q_3$ is in fact
obtained by intersecting the elements corresponding
to $q_1$ and $q_2$.
\be
q_3\nn=q_1\nn\cap q_2\nn\ \ .
\ee
The $\Leftarrow$ part can be proven as follows. Suppose $q_3\preceq q_1$,
this implies that, at level $n$ we have that
\be
q\nn_3\subseteq q\nn_1\ \ ,
\ee
 but then
\be
\{\pi(q_3)\}=\bigcap_n q_3\nn \subseteq \bigcap_n q_1\nn =
\{\pi(q_1)\}\ \ .
\ee
{}From this equation it follows that $\pi(q_3)=\pi(q_1)$, hence $q_3\sim q_1$.
One can prove $q_3\sim q_2$ in an analogous way.
\end{itemize}

So we now have a characterization of $P\inff/\sim$ which is
independent of the
projection $\pi$, in other words $\sim$ is uniquely defined once
$P\inff$ is given.
With the use of noncommutative
lattices we see then that, on one side, the
lattice at finite level has a memory of the topology of the original manifold,
on the other side, in the limit it is possible to recover
unambiguously the original
(Hausdorff) manifold $M$.

Before going to the algebraic description of the noncommutative lattices and
their
continuum limit let us consider a
construction which will be useful in the following. A simple way to
obtain the noncommutative lattice for a topological space, which is the
cartesian product
of lower dimensional spaces, is through the cartesian product of the
corresponding noncommutative lattices.
Suppose that $P$ is the cartesian product of the noncommutative lattices $P^
i$'s. A
relation on $P$ is given by declaring that any
two elements of $P$ are related iff each component of one element
is related to the corresponding component of the other.
As an example, consider the cartesian product of a noncommutative lattice
$P = \{p_1, \cdots, p_n \}$ with a noncommutative lattice $Q = \{q_1, \cdots,
q_m \}$.
The cartesian product
\be
P \times Q = \{(p_j, q_k)~~p_j \in P~, q_k \in Q\}~,
\ee
is given the order relation
\be
(p_i, q_j) \preceq (p_k, q_l)
\iff p_i \preceq p_k ~~{\rm and}~~q_j \preceq q_l~.
\ee
A simple example is shown in fig. 2.4

\begin{figure}[htb]
\begin{center}
\unitlength=1.00mm
\special{em:linewidth 0.4pt}
\linethickness{0.4pt}
\begin{picture}(48.00,63.00)
\put(17.00,20.00){\circle*{2.00}}
\put(47.00,20.00){\circle*{2.00}}
\put(5.00,32.00){\circle*{2.00}}
\put(5.00,62.00){\circle*{2.00}}
\put(17.00,32.00){\circle*{2.00}}
\put(17.00,62.00){\circle*{2.00}}
\put(47.00,62.00){\circle*{2.00}}
\put(47.00,32.00){\circle*{2.00}}
\put(32.00,46.00){\makebox(0,0)[cc]{$\diamondsuit $}}
\put(21.00,62.00){\line(1,0){7.00}}
\put(36.00,62.00){\line(1,0){7.00}}
\put(21.00,47.00){\line(1,0){7.00}}
\put(36.00,47.00){\line(1,0){7.00}}
\put(21.00,32.00){\line(1,0){7.00}}
\put(36.00,32.00){\line(1,0){7.00}}
\put(21.00,20.00){\line(1,0){7.00}}
\put(36.00,20.00){\line(1,0){7.00}}
\put(5.00,58.00){\line(0,-1){7.00}}
\put(5.00,43.00){\line(0,-1){7.00}}
\put(17.00,58.00){\line(0,-1){7.00}}
\put(17.00,43.00){\line(0,-1){7.00}}
\put(32.00,58.00){\line(0,-1){7.00}}
\put(32.00,43.00){\line(0,-1){7.00}}
\put(47.00,58.00){\line(0,-1){7.00}}
\put(47.00,43.00){\line(0,-1){7.00}}
\put(32.00,62.00){\makebox(0,0)[cc]{$\times $}}
\put(5.00,47.00){\makebox(0,0)[cc]{$\times $}}
\put(17.00,47.00){\makebox(0,0)[cc]{$\times $}}
\put(47.00,47.00){\makebox(0,0)[cc]{$\times $}}
\put(32.00,32.00){\makebox(0,0)[cc]{$\times $}}
\put(32.00,20.00){\makebox(0,0)[cc]{$\times $}}
\end{picture}
\end{center}
{\footnotesize {\bf Fig. 2.4.} The figure shows the cartesian product
$P^1\times P^1$. Points connected by links are ordered according to the
convention $\bullet \preceq \times \preceq \diamondsuit $. }
\end{figure}

With this construction one can obtain the noncommutative lattice for any
$d$-dimensional cubic  decomposition of a space $M$.
One takes the cartesian product of $d$ copies of the noncommutative lattice
$P^1(I)$
(like the 2-dimensional example of fig. 2.4)
for each elementary $d$-dimensional cube in the decomposition.
One then joins the cubes by identifying some of the faces.
The
projective system $\{P^n(M), \pi^{(n,m)}(M)\}$
and the projective limit $P^\infty (M)$ space can also be
obtained in this way from the corresponding ones of the interval.
If $M$ is a
$d$-dimensional cube $C_d$, a projective system is given by
$\{P^n(C_d), \pi^{(n,m)}(C_d)\}$ where
\be
P^n(C_d) = \underbrace{P^n\times P^n\times\dots\times P^n}_d~,
\label{prod}
\ee
where $P^n$ are the noncommutative lattices of the interval, and
\be
\pi^{(m,n)}(C_d)(p_1,\cdots, p_d)\equiv (\pi^{(m,n)}p_1,\cdots,
\pi^{(m,n)}p_d).
\ee
The projective limit is
\be
P^\infty(C_d)  = \underbrace{P^\infty\times
P^\infty\times\dots\times P^\infty}_d.
\ee
Finally the cube itself is obtained from $P^\infty(C_d)$ as the
quotient defined by the projection $\pi(C_d) = (\pi,\cdots,\pi)$ from
$P^\infty(C_d)$ onto the cube.

This construction has a straightforward algebraic translation in
terms of tensor products.

\sxn{The Continuum Limit of Algebras}\label{se:cla}

\subsxn{The Algebras of Noncommutative Lattices}

We shall now associate with each noncommutative lattice $P$ an algebra of
operator valued functions which contain the same topological information, in
the
sense that $P$ can be reconstructed as a topological space from the
algebra.
For a Hausdorff space $M$, the natural choice would be the algebra of
continuous functions ${\cal C}(M)$ on $M$. As it is known \cite{FD}, from
${\cal C}(M)$ the space $M$ itself can be
reconstructed as the set ${\widehat{{\cal C}(M)}}$ of irreducible
representations (IR's) of ${\cal C}(M)$ or,
equivalently, as the set of complex homomorphisms of ${\cal C}(M)$. The
topology on ${\widehat{{\cal C}(M)}}$ is then defined by pointwise convergence,
\be
p_n\rightarrow p ~~\Longleftrightarrow~~ p_n(f)\rightarrow p(f) ~~~
\forall f\in {\cal C}(M), ~~p_n, p\in{\widehat{{\cal C}(M)}}.
\ee
Actually there is a complete
correspondence between abelian $C^*$-algebras (with unit) and
(compact) Hausdorff spaces \cite{FD}.

However,
since our noncommutative lattice $P$ is not Hausdorff, the usual algebra of
${\complex }$-valued  continuous
functions on $P$ is not able to capture the topological
structure of $P$. It is easy to see, indeed, that the only continuous
complex
valued functions on $P$ are the constant ones and
${\cal C}(P)$ is then equal to
${\complex }$. Since $\widehat{\complex } = \widehat{{\cal C}(P)}$
consists of a single
point, we see that the structure space of ${{\cal C}(P)}$ is not $P$ and is
actually trivial. The algebra ${\cal C}(P)$, in other words, because of the
continuity property of its elements, identifies those points of
$P$ which cannot be separated by the topology and thus gives rise to
a space with a single point.

It is however possible to reconstruct the noncommutative lattice $P$ and its
topology from a {\it noncommutative} \cstar of operator-valued functions.
Given a noncommutative $C^*$-algebra, $\ca$, one can construct a
topological space $\widehat \ca$, not Hausdorff in general, by considering
again the set of all IR's (in general not anymore one dimensional)
of $\ca$.
One gives a topology to $\widehat \ca$, called regional
topology, by giving a notion of closeness among IR's of
different dimensions \cite{FD}. For a particular class of algebras,
called postliminal algebras \cite{FD}, which includes the ones we will
be interested in, there is an equivalent but easier,
construction of $\widehat \ca$ in terms of a particular class of ideals
of $\ca$. The kernels of IR's of an algebra $\ca$ are
bilateral ideals, called primitive ideals. Generally, different IR's
may have the same kernel, however for postliminal algebras there is a
one to one correspondence between IR's and primitive ideals. For
these algebras then, the space $\widehat \ca$ can be equivalently regarded
as the set of primitive ideals and the topology can be equivalently
given  as the Hull kernel topology. This topology is defined by
giving the procedure to construct the closure, ${\bar S}$, of an
arbitrary set, $S$, of primitive ideals. If $S = \{\pr_\lambda, \
\lambda\in \Lambda \}$ where $\pr_\lambda$ is a family of primitive ideals
parametrized by
$\lambda$, then
\be
ker S \equiv  \{\bigcap_{\lambda\in\Lambda}\pr_\lambda\}
\ee
and
\be
{\bar S} \equiv Hull~ker S \equiv \{\pr : ker S \subset \pr, \
\pr\in \widehat{\ca}\}.
\ee

Since what we have said about primitive ideals holds also in the
commutative case, we give a simple abelian example to clarify this definition.
Consider the
algebra $C(I)$ of complex continuous functions on an interval $I$, and
for $a<b \in I$ let $S$ be
\be
S = \{\pr_\lambda, \ \lambda \in ]a,b[\}
\ee where $\pr_\lambda$ is the primitive ideal of $C(I)$ given by the
kernel of the homomorphism
\be
\lambda : f\rightarrow \lambda(f)\equiv f(\lambda) \ \ \ f\in C(I)
\ee
that is the set of functions vanishing at $\lambda$
\be
\pr_\lambda = \{f : f(\lambda)= 0 \}.
\ee
$Ker S$ is then
\be
ker S = \bigcap_{\lambda\in ]a,b[}\pr_\lambda =\{f : f(\lambda) = 0 \
\forall \lambda\in ]a,b[\}.
\ee
Now, since the functions $f$ are continuous, we also have
\bea
{\bar S}= Hull~ker S &=& \{\pr : ker S\subset \pr\}  \\
                    &=& S~\bigcup~ \{\pr_a,\pr_b\}    \\
                    &=& \{\pr_\lambda,~ \lambda\in [a,b] \}.
\eea
which IS the closure of the open interval $]a,b[$.

{}From this definition of topology we see that if an
ideal $\pr_{\lambda_1}$ is included in an ideal $\pr_{\lambda_2}$,
then the point $\pr_{\lambda_2}$, being in the closure of
$\pr_{\lambda_1}$, is a limit point of $\pr_{\lambda_1}$. The
topology of a noncommutative lattice or, which is the same,
the order relation of the poset, is
then simply encoded in the inclusion relation among the primitive
ideals of the corresponding algebra.

As it is proven in  \cite{FD}, each finite $T_0$
topological space is the structure space of a noncommutative
$C^*$-algebra.

Before giving few examples of the
reconstruction theorem we recall few facts
about compact operators which play a crucial role in our algebras.
An operator in a
Hilbert space $\ch$ is said to be of finite rank if the orthogonal
complement
of its null space is finite dimensional. An operator $k$ in
$\ch$ said to be compact if it can be approximated arbitrarily
closely in norm
by finite rank operators. If $\l_1,\l_2,$ ... are the eigenvalues of
$k^* k$ for such a $k$, with $\l_{i+1} \leq \l_i$ and
an eigenvalue of multiplicity $n$
occurring $n$ times in this sequence, (here and in what follows, $*$
denotes the adjoint for an operator)
then $\l_n\rightarrow 0$ as
$n\rightarrow \infty $. It follows that the operator $1\!\!1$ in an infinite
dimensional Hilbert space is not compact.

The set $\ck$ of all compact operators $k$ in a Hilbert space is a \cstar.
It is a two-sided ideal in the \cstar $\cb$ of all bounded operators
\cite{FD}.
The construction of $\ca$ for a noncommutative lattice rests on the
following result from the
representation theory of $\ck$. The representation of $\ck$ by itself is
irreducible \cite{FD} and it is the \underline{only} IR of $\ck$ up to
equivalence.

\bigskip

\noindent{\bf Example 3.1.}
The simplest nontrivial noncommutative lattice is
$\widetilde{P} = \{p_1, p_2 \}$ with $p_1
\prec p_2$. It is
shown in fig. 3.1 (a).

\begin{figure}[htb]
\begin{center}
\unitlength=1.00mm
\special{em:linewidth 0.4pt}
\linethickness{0.4pt}
\begin{picture}(61.00,45.00)
\put(0.00,20.00){\circle*{2.00}}
\put(7.00,20.00){\makebox(0,0)[cc]{$p_1$}}
\put(7.00,45.00){\makebox(0,0)[cc]{$p_2$}}
\put(50.00,20.00){\circle*{2.00}}
\put(57.00,20.00){\makebox(0,0)[lc]{$a(p_1)=\lambda \I + k$}}
\put(57.00,45.00){\makebox(0,0)[lc]{$a(p_2)=\lambda $}}
\put(4.00,7.00){\makebox(0,0)[cc]{(a)}}
\put(61.00,7.00){\makebox(0,0)[cc]{(b)}}
\put(0.00,41.00){\line(0,-1){17.00}}
\put(50.00,41.00){\line(0,-1){17.00}}
\put(50.00,45.00){\makebox(0,0)[cc]{$\times $}}
\put(0.00,45.00){\makebox(0,0)[cc]{$\times $}}
\end{picture}
\end{center}
{\footnotesize {\bf Fig. 3.1.} (a) is the poset $\widetilde{P}$
with two points. (b)
shows the values of a generic element $\lambda \I + k$ of its algebra
{\ca} at its two points $p_1$ and $p_2$.}
\end{figure}

We associate an infinite dimensional Hilbert space $\ch$ to this
poset. The corresponding algebra is then the subalgebra of the
bounded operators $\cb(\ch)$ given by
\be
\ca= {\complex }1\!\!1 + \ck(\ch) =
             \{ \l 1\!\!1 + k \, : \, \l\in {\complex }, k \in
\ck(\ch) \}
\;.\label{4.1}
\ee

In order to construct the IR's of $\ca$ we recall a known theorem
\cite{Mu} which states that the IR's of a
$C^*-subalgebra$ of ${\cb}(\ch)$ which includes the algebra of
compact operators $\ck(\ch)$ are of two kinds: either they vanish
on the compact operators or they are unitary equivalent to the
defining representation. In our case then we will have only two IR's
(up to unitary equivalence), the first one, $\pi_1$, which is one
dimensional and vanishes on $\ck(\ch)$, is given by
\be
\pi_1(\lambda\I_{\ch} + k) = \lambda.
\ee
The second one, $\pi_2$, is just the defining representation:
\be
\pi_2(a) = a , \ a\in \ca.
\ee
The kernels ${\cal I}_1$, ${\cal I}_2$ of $\pi_1$ and $\pi_2$ are
then the only two primitive ideals of $\ca$, and are
\bea
{\cal I}_1 &=& \ck(\ch) \\
{\cal I}_2 &=& \{0\}.
\eea
Since the second ideal is included in the first one we see that
$\widehat \ca$
is actually the noncommutative lattice
$\widetilde P$ of fig. 3.1 (a)
if we identify ${\cal I}_1$ and
${\cal I}_2$ with $p_1$ and $p_2$ respectively.
An arbitrary element $\l 1\!\!1 +k$ of $\ca$ can be regarded as a ``function"
on it if, in analogy to the commutative case, we set
\bea
(\l 1\!\!1 +k) (p_2) &:=& \l \; \nonumber \\
(\l 1\!\!1 +k) (p_1) &:=& \l 1\!\!1 +k \; . \label{4.1b}
\eea

Notice that in this case the function $\l 1\!\!1 +k$ is not valued
in ${\complex }$ at all points. Indeed, at different points it is valued
in different spaces, ${\complex }$ at $p_2$ and a subset of bounded operators
on an infinite Hilbert space at $p_1$.

\bigskip

\noindent
{\bf Example 3.2.}
We next consider the noncommutative lattice $V$ consisting of three points
$p_1, p_2, p_3$ as in fig. 3.2.

\begin{figure}[htb]
\begin{center}
\unitlength=1.00mm
\special{em:linewidth 0.4pt}
\linethickness{0.4pt}
\begin{picture}(49.00,50.00)
\put(-8.00,35.00){\circle*{2.00}}
\put(-8.00,46.00){\line(0,-1){7.00}}
\put(-8.00,31.00){\line(0,-1){7.00}}
\put(-1.00,20.00){\makebox(0,0)[cc]{$p_1$}}
\put(-1.00,35.00){\makebox(0,0)[cc]{$p_3$}}
\put(-1.00,50.00){\makebox(0,0)[cc]{$p_2$}}
\put(-15.00,43.00){\makebox(0,0)[cc]{$\ch_2$}}
\put(-15.00,28.00){\makebox(0,0)[cc]{$\ch_1$}}
\put(42.00,35.00){\circle*{2.00}}
\put(42.00,46.00){\line(0,-1){7.00}}
\put(42.00,31.00){\line(0,-1){7.00}}
\put(49.00,50.00){\makebox(0,0)[lc]{$a(p_2)=\lambda _2$}}
\put(49.00,35.00){\makebox(0,0)[lc]{$a(p_3)=\lambda _1{\cal P}_1 +
\lambda _2{\cal P}_2 + k$}}
\put(49.00,20.00){\makebox(0,0)[lc]{$a(p_1)=\lambda _1$}}
\put(49.00,10.00){\makebox(0,0)[cc]{(b)}}
\put(-8.00,10.00){\makebox(0,0)[cc]{$(a)$}}
\put(42.00,50.00){\makebox(0,0)[cc]{$\times $}}
\put(-8.00,50.00){\makebox(0,0)[cc]{$\times $}}
\put(-8.00,20.00){\makebox(0,0)[cc]{$\times $}}
\put(42.00,20.00){\makebox(0,0)[cc]{$\times $}}
\end{picture}
\end{center}
{\footnotesize {\bf Fig. 3.2.} (a) shows the noncommutative lattice V and
the
association of an infinite dimensional Hilbert space ${\cal H}_i$ to
each of its "arms''. (b) shows the values of a typical element
$a=\lambda _1{\cal P}_1 + \lambda _2{\cal P}_2 + k$ of its algebra at
its three points.}
\end{figure}

The noncommutative lattice $V$ has two arms 1 and 2.
In order to construct the associated algebra,
we attach an infinite-dimensional Hilbert space $\ch_i$ to each arm $i$ as
shown in fig. 3.2.
Let $\cp_i$ be the orthogonal projector on $\ch_i$ in $\ch_1\oplus
\ch_2$ and $\ck_{12}=\{k_{12}\}$ be the set of all compact operators in
$\ch_1\oplus \ch_2$. The associated algebra is then
\be
        \ca= {\complex }\cp_1 + {\complex }\cp_2 + \ck_{12} ~ .\label{4.2}
\ee
We will indicate an element $a$ of this algebra either as $a = \l_1\cp_1
+ \l_2\cp_2 + k$ or, with an obvious notation, $a = (\l_1,\l_2) + k$.
Using the same characterization of IR's as in the
previous example we can see that this algebra has three IR's $~\pi_1,
\pi_2, \pi_3$. The representations
$\pi_1$ and $\pi_2$, which are the ones vanishing on the
compact operators, are one-dimensional and are given by
\bea
\pi_1(\lambda_1 \cp_1 + \lambda_2 \cp_2 + k) &=& \lambda_1  \\
\pi_2(\lambda_1 \cp_1 + \lambda_2 \cp_2 + k) &=& \lambda_2 .
\eea
The representation
$\pi_3$ is the defining one of $\ca$:
\be
\pi_3(a) = a \ \ \ a\in \ca.
\ee
The corresponding kernels, $\pr_1,\pr_2,\pr_3$ are
\bea
\pr_1 &=& ker \pi_1 = \{ a\in \ca : a = \lambda \cp_2 + k\}  \\
\pr_2 &=& ker \pi_2 = \{ a\in \ca : a = \lambda \cp_1 + k\}  \\
\pr_3 &=& ker \pi_3 = \{0\}.
\eea
Since  the inclusion relations among the primitive ideals are
$\pr_3\subset \pr_1$ and $\pr_3\subset \pr_2$, we see that, with the
identifications $p_i = \pi_i$, $\widehat \ca$ is equal to $P$.

\bigskip

\noindent
{\bf Example 3.3.}
In general,
consider a noncommutative lattice $P$ which is obtained by joining two other
noncommutative lattices $P_1$ and
$P_2$ by identifying some closed set of points. This operation has a
simple algebraic
translation. The algebra $\ca$ of $P$ is obtained in two steps; first
consider the direct sum ${\bar \ca} = \ca_1\oplus \ca_2$ of the two algebras
associated to $P_1$ and $P_2$. The algebra $\ca$ of $P$
is then equal to the subalgebra of
${\bar \ca}$ consisting of those elements which have the same value at the
points (seen as IR's) which are identified.

Consider the construction of the algebra $\ca_n$ of a
generic noncommutative lattice $P^n$ for the interval.
Such a noncommutative lattice is composed of
a number $N = 2^{n-1}-1$ of $V$'s plus the elementary
noncommutative lattice ${\widetilde P}$ with
two points, of example 3.1, at each end.
Number the arms and attach an
infinite dimensional Hilbert space $\ch_i$ to each arm $i$.
To a $V$ with arms
$i,i+1$, attach the algebra $\ca_i$ with elements
$\l_i\cp_i+\l_{i+1}\cp_{i+1}+k_{i,i+1}$. Here $\l_{i},\l_{i+1}$ are any two
complex numbers, $\cp_i ,\cp_{i+1}$
are orthogonal projectors on $\ch_i$, $\ch_{i+1}$ in the Hilbert
space $\ch_i\oplus \ch_{i+1}$ and $k_{i,i+1}$ is any compact operator
in $\ch_i\oplus \ch_{i+1}$. This is as before. But now, for gluing
the various algebras together, we also impose the
condition $\l_j=\l_k$ if the lines $j$ and $k$ meet at a top point.
Taking into account also the algebras $\ca_0$ and $\ca_{N+1}$ for the
noncommutative lattice ${\tilde P}$
at the beginning and at the end, we can express the
algebra $\ca_n$ as a direct sum  plus this
condition. A generic element $a_n$ of $\ca_n$ has then the form
\be
        a= (\l_1\cp_1 + k_{1}) \bigoplus_{i=1}^N a_i
        \oplus (\l_{N+1}\cp_{2N+2} + k_{2N+2}), ~~~~~
a_i = \l_i\cp_{2i}+\l_{i+1}\cp_{2i+1}+k_{2i,2i+1} \label{4.3}
\ee

\begin{figure}[htb]
\begin{center}
\unitlength=1.00mm
\special{em:linewidth 0.4pt}
\linethickness{0.4pt}
\begin{picture}(146.00,45.00)
\put(115.00,30.00){\circle*{2.00}}
\put(145.00,30.00){\circle*{2.00}}
\put(89.00,30.00){\line(1,0){6.00}}
\put(103.00,30.00){\line(1,0){8.00}}
\put(119.00,30.00){\line(1,0){6.00}}
\put(133.00,30.00){\line(1,0){8.00}}
\put(5.00,30.00){\circle*{2.00}}
\put(35.00,30.00){\circle*{2.00}}
\put(9.00,30.00){\line(1,0){6.00}}
\put(23.00,30.00){\line(1,0){8.00}}
\put(39.00,30.00){\line(1,0){6.00}}
\put(53.00,30.00){\line(1,0){8.00}}
\put(75.00,30.00){\makebox(0,0)[cc]{. . .}}
\put(18.00,40.00){\line(1,0){32.00}}
\put(50.00,40.00){\line(0,-1){5.00}}
\put(18.00,40.00){\line(0,-1){5.00}}
\put(98.00,40.00){\line(1,0){32.00}}
\put(130.00,40.00){\line(0,-1){5.00}}
\put(98.00,40.00){\line(0,-1){5.00}}
\put(115.00,45.00){\makebox(0,0)[cc]{V}}
\put(35.00,45.00){\makebox(0,0)[cc]{V}}
\put(49.00,30.00){\makebox(0,0)[cc]{$\times $}}
\put(19.00,30.00){\makebox(0,0)[cc]{$\times $}}
\put(99.00,30.00){\makebox(0,0)[cc]{$\times $}}
\put(129.00,30.00){\makebox(0,0)[cc]{$\times $}}
\end{picture}
\end{center}
{\footnotesize {\bf Fig. 3.3.} A noncommutative lattice for the segment
constructed by gluing several V posets and a pair of two point posets at the
ends.}
\end{figure}

There is a systematic and simple construction of $\ca$ for any
noncommutative
lattice (that is, any ``finite $T_0$ topological space") which is given in
\cite{BL}.

It should be remarked that actually
the noncommutative lattice does not uniquely fix its algebra as there are in
general many
non-isomorphic (noncommutative) C$^*$-algebras with the same poset
as structure space \cite{BL}. This is to be contrasted with
the Gel'fand-Naimark result asserting that the (commutative)
\cstar associated to a
Hausdorff topological space (such as a manifold) is unique.

\subsxn{Direct Systems of Algebras and their Limits}

In this section we shall consider the question of the continuum limit
of the system of algebras $\ca_n$ associated with the
noncommutative lattices $P^n$
which are
obtained from the cubic decompositions of a topological space $M$.
We shall construct an inductive system of algebras
whose inductive limit $\ca_\infty$ will have the following
two properties:
\begin{itemize}
\item[a)] The structure space of $\ca_\infty$ is $P^\infty$.
\item[b)] The algebra of continuous functions on $M$ coincides with
       the  center of $\ca_\infty.$
\end{itemize}

We prove these statements for the interval $I$. We shall then show
how to extend the results to any topological space
$M$ admitting cubic decompositions.

A natural framework to define the continuum limit of the
algebras $\ca_n$ is that of inductive system of algebras which dualizes
the projective system of noncommutative lattices. Let us then recall the
notion of
inductive system and inductive limit of $C^*$-algebras.
An {\it inductive system} of $C^*$-algebras is a sequence of $C^*$-algebras
$\ca_n$, together with norm non-increasing immersions
$\Phi_{(n,m)}~:~\ca_n \rightarrow \ca_m,~n<m$, such that the
composition law $\Phi_{{(n,m)}}\Phi_{{(m,p)}}=\Phi_{{(n,p)}},~n<m<p~,$ holds.

The inductive limit $\ca_\infty$ is the \cstar consisting
of equivalence classes
of ``Cauchy sequences" $\{a_n \}, \ \ a_n \in \ca_n$. Here by Cauchy
sequence we mean that $||\Phi_{{(n,m)}}(a_n) -a_m||_m$ goes to zero
as $n$ and $m$ go to infinity.
Two sequences $\{a_n\}$ and $\{b_n\}$ are equivalent if
$||a_n - b_n||_{n}$ goes to zero. The norm in $\ca_\infty$ is defined by
\be
||a||_\infty = \lim_{n\rightarrow\infty} ||a_n||_{n}
\ee
where $\{a_n\}$ is any of the representatives of $a$.
In case the algebras $\ca_n$ are realized as an increasing sequence of
$C^*$-subalgebras of a given $C^*$-algebra $\ca$, the inductive limit
is just the closure of the union of the algebras $\ca_n$.
\fn{For a more
detailed account of the definition see for example \cite{FD} or
\cite{Mu}.}.

Let us now consider the interval $I = [0,1]$. Its projective system
of noncommutative lattices $P^n$ has been described in section 2 and is
illustrated
in fig. 2.2. The algebra $\ca_n$ of $P^n$ has been introduced in
section 3.1 and the form of the generic element is given in eq.
(\ref{4.3}).

In this case the immersions $\Phi_{(n,m)}$ are obtained through a suitable
definition of pull-back of the projections $\pi^{(m,n)}$. The
usual definition of pull-back would be
\be
p^m_{l\alpha}(\Phi_{(n,m)}(a^n)) = p^n_{l'\alpha'}(a_n) \label{pb1}
\ee
where
\be
\pi^{(m,n)}(p^m_{l\alpha}) = p^n_{l'\alpha'},~~~~~~~~a^n\in \ca_n,
{}~~p^m_{l\alpha}\in P^m, ~~p^n_{l'\alpha'}\in P^n.
\ee
(Here we are thinking of the elements $p^n_{l \alpha} \in P^n$
as $IR$'s of $\ca_n$.)
Eq. (\ref{pb1}) however, makes sense only when the
two representations $p^m_{l\alpha} $ and $ p^n_{l'\alpha'}$
have the same dimension but this is not always the case. Due to
the refinement procedure of
the decompositions and to the properties of the
projections $\pi^{(m,n)}$, the representations have different
dimensions when $l = 0$ and
$l' = 1$. Then, $p^m_{0\alpha} $ is a point
of $P^m$ corresponding to a zero dimensional cube and thus is an
infinite dimensional
representation
\be
p^m_{0\alpha}(a^m) = \mu_\alpha \I_{\ch_\alpha} +
\mu_{\alpha+1}\I_{\ch_{\alpha+1}} + k_\alpha~.
\ee
Also, $ p^n_{1'\alpha'}$ is a point of $P^n$ corresponding to a one
dimensional cube and is the one dimensional representation
\be
p^n_{1\alpha'}(a^n) = \lambda_{\alpha'}.
\ee
Eq. (\ref{pb1}) is replaced in this case by
\be
p^m_{0\alpha}(\Phi_{(n,m)}(a^n)) = p^n_{1\alpha'}(a_n)\otimes
\I_{{\ch_\alpha} \oplus \ch_{\alpha+1}}.
\ee
This equation is now meaningful and can be solved for the  value of
$\Phi_{(n,m)}(a^n)$ at the point $p^m_{0\alpha}$. More explicitly, it gives
\be
\mu_\alpha \I_{\ch_\alpha} + \mu_{\alpha+1}\I_{\ch_{\alpha+1}} + k_\alpha =
\lambda_{\alpha'}\otimes \I_{{\ch_\alpha} \oplus \ch_{\alpha+1}}
\ee
whose solution is
\be
\mu_\alpha = \mu_{\alpha+1} = \lambda_{\alpha'}, ~~~~~k_\alpha = 0~.
\ee
This is shown in fig. 3.4. In general, for $\Phi_{(n,n+1)} : \ca_n
\rightarrow \ca_{n+1} $ we have
\bea
\Phi_{n,n+1}[(\l_1\cp_1 + k_1 ) \bigoplus_{i=1}^N
   (\l_i\cp_{2i}+\l_{i+1}\cp_{2i+1}+k_{2i,2i+1})
  \oplus (\l_{N+1}\cp_{2N+2} + k_{2N+2})] = \nonumber \\
= (\l_1\cp_1 + k_1) \bigoplus_{i=1}^N
\left[ (\l_i\cp_{4i-2} + \l_i \cp_{4i-1})
\oplus (\l_{i}\cp_{4i}+\l_{i+1}\cp_{4i+1}+k'_{4i,4i+1})\oplus
\right.\nonumber \\
\oplus
\left. (\l_{i+1}\cp_{4i+2}+ \l_{i+1} \cp_{4i+3}) \right]
\oplus (\l_{N+1}\cp_{4N+4} + k'_{4N+4})
\eea
where $k'_{4i,4i+1}=k_{2i,2i+1}$ and $k'_{4N+4}=k_{2N+2}$.

\begin{figure}[htb]
\begin{center}
\unitlength=1.00mm
\special{em:linewidth 0.4pt}
\linethickness{0.4pt}
\begin{picture}(86.00,49.00)
\put(5.00,20.00){\circle*{2.00}}
\put(5.00,40.00){\circle*{2.00}}
\put(45.00,40.00){\circle*{2.00}}
\put(85.00,40.00){\circle*{2.00}}
\put(85.00,20.00){\circle*{2.00}}
\put(41.00,20.00){\line(-1,0){32.00}}
\put(49.00,20.00){\line(1,0){32.00}}
\put(81.00,40.00){\line(-1,0){12.00}}
\put(61.00,40.00){\line(-1,0){12.00}}
\put(41.00,40.00){\line(-1,0){12.00}}
\put(21.00,40.00){\line(-1,0){12.00}}
\put(5.00,36.00){\vector(0,-1){12.00}}
\put(45.00,36.00){\vector(0,-1){12.00}}
\put(85.00,36.00){\vector(0,-1){12.00}}
\put(65.00,36.00){\vector(-4,-3){16.00}}
\put(25.00,36.00){\vector(4,-3){16.00}}
\put(85.00,49.00){\makebox(0,0)[cc]{$\lambda \I + k_2$}}
\put(65.00,49.00){\makebox(0,0)[cc]{$\lambda $}}
\put(45.00,49.00){\makebox(0,0)[cc]{$\lambda \I$}}
\put(25.00,49.00){\makebox(0,0)[cc]{$\lambda $}}
\put(5.00,49.00){\makebox(0,0)[cc]{$\lambda \I + k_1$}}
\put(5.00,10.00){\makebox(0,0)[cc]{$\lambda \I +k_1$}}
\put(45.00,10.00){\makebox(0,0)[cc]{$\lambda $}}
\put(85.00,10.00){\makebox(0,0)[cc]{$\lambda \I + k_2$}}
\put(25.00,40.00){\makebox(0,0)[cc]{$\times $}}
\put(65.00,40.00){\makebox(0,0)[cc]{$\times $}}
\put(45.00,20.00){\makebox(0,0)[cc]{$\times $}}
\end{picture}
\end{center}
{\footnotesize {\bf Fig. 3.4.} The figure shows two noncommutative lattices
corresponding to two consecutive levels of approximation. The arrows
indicates the projection. The pullback of a generic function on the lower
level to the upper level is indicated. }
\end{figure}

With these immersions $\Phi_{(n,m)}$, the set of algebras $\ca_n$ is made
into an inductive system.
We want to find now the inductive limit $\ca_\infty$, and prove that
it is the dual of $P^\infty$ and that its center is $C(I)$.

Instead of constructing abstractly the inductive limit $\ca_\infty$ we
will represent it as a $C^*$-algebra of operator valued functions on the
interval. This construction is suggested by the following theorem due
to Dauns and Hofmann \cite{DB}:

{\it Dauns-Hofmann theorem:} let $\ca$ be a $C^*$-algebra with identity
and $M$ the space of maximal
ideals of the center of $\ca$ with the Hull-kernel topology. Then
$\ca$
is isometrically $^*$-isomorphic to the $C^*$-algebra of all
continuous sections $\Gamma(\pi)$ of a $C^*$- bundle $\xi =
(\pi,B,M)$ over $M$. The fiber above $m\in M$ is the quotient
$C^*$-algebra $\ca / m\ca$, the isometric $^*$-isomorphism is the
Gel'fand representation $x\rightarrow {\hat x}$, where ${\hat x}(m) = x
+m\ca $, and the norm of ${\hat x}$ is given by
\be
||{\hat x}|| =sup\{||{\hat x}(m)|| : m\in M\}.
\ee
Further, the real valued map $m\rightarrow ||{\hat x(m)}||$ on $M$ is
upper semicontinuous \cite{EDM} for each $x\in \ca$.

Let us then construct the $C^*$-algebra $\ca_\infty$ as an
algebra of operator valued functions on $I$.
Recall that $P^\infty$ is a quasi fiber bundle on $I$
whose fibers at the points of the set $I_0 = \{x =
\bigcup_{n,\alpha}S^n_{0,\alpha}\}$  are the  noncommutative lattice $V$ (
see
example 2.2) while they are made of a single point for $I - I_0$. Now, the
algebra corresponding to $V$ is $\{\lambda_1\I_{\ch_1} +
\lambda_2\I_{\ch_2} + k\}$
while the one corresponding to a point is ${\complex }$. We are thus led to
consider an element of $\ca_\infty$ as an operator valued function on
$I$ of the kind
\be
a(x) =
\left\{
\begin{array}{cc}
(\lambda_+(x), \lambda_-(x)) + k(x)  &x\in I_0 \\
 \lambda(x)~~~~~~~~~~~~~~~~~~~~~~~~&x\in I- I_0
\end{array}
\right.
\ee
where :
\begin{itemize}
\item[a)] the function $\lambda(x)$ is continuous and
bounded in $I - I_0$ and $\lim_{x\rightarrow {\bar x}^{\pm}}\lambda(x)
= \lambda_{\pm}({\bar x})$ for all ${\bar x} \in I_0$.

\item[b)] $\forall \epsilon\geq 0$ there exist only a finite number of
points
${\bar x}\in I_0$ such that

$|| (\lambda_+({\bar x}), \lambda_-({\bar x})) + k({\bar x})||> \epsilon $.
\end{itemize}
The norm of any $a(x) \in \ca_\infty$ is defined as
$||a|| = sup_{x \in I} ||a(x)||$.

It is easy to see that any element $a_n$ at finite level can be regarded as a
function of this kind having as $\lambda(x)$ a stepwise function and
$k(x)$
different from zero only at a finite number of points. This provides an
embedding $\Phi_n : \ca_n \rightarrow \ca_{\infty}$. It can then be verified
that, due to conditions $a$ and $b$,  the  closure of the union of the algebras
$\Phi_n (\ca_n)$ is $\ca_\infty$ so that $\ca_\infty$ is the inductive limit of
the system $\{\ca_n,\pi^{(m,n)}\}$.

Although we have given the structure of the fibers and realized the
algebra $\ca_\infty$ as sections, we have not specified yet
the topology of the bundle with respect to which
$\ca_\infty$ is the set of all continuous functions.
The existence of such a topology, however, follows from the fact that
our presentation of $\ca_\infty$ coincides with the
one given in the Dauns-Hofmann theorem.
Indeed,
\begin{itemize}
\item[1.]
The center of $\ca_\infty$ is the set of continuous
functions of the interval, namely is the set of sections such that
\be
\lambda_-(x) = \lambda_+(x)~,~~~ k(x) = 0~, ~~~~~ \forall x \in I.
\label{center}
\ee
Indeed, since the product is pointwise, even the characterization of
the center can be done pointwise. Now, only at the points $x$ in
$I_0$ the algebras $\ca_\infty(x)$ will be nonabelian, and their
elements will be of the form $(\lambda_-(x),\lambda_+(x)) + k(x)$.
It is then easy to check that for an
element of this kind to be in the center,
eq.(\ref{center}) has to be satisfied at that point.

\item[2.]
One easily proves that the fibers, $\ca(x)$, over any $x \in I$,
given by the Dauns-Hofmann theorem
as the quotient $\ca_\infty / x \ca_\infty$,
coincides with ours. They are either a copy of ${\complex }$ if
$x \in I - I_0$ or a copy of the algebra for a $V$ if $x \in I_0$.

\item[3.]
The sections that, according to the theorem,
should be associated to elements of $\ca_\infty$ are the sections
that we used to define the elements themselves.

\end{itemize}

\bigskip

Finally, we want to show that the structure space of $\ca_\infty$ is
just $P^\infty$.
Now, under the conditions of the Dauns-Hofmann theorem, by a
result due to Varela \cite{Va}, each IR of
$\ca_\infty$ factors through the evaluation map, namely, given an IR
$~T$ of $\ca_\infty$ there exists a unique point $x\in I$ and a point
$q\in {\widehat \ca}(x)$ such that
\be
T(a) = q(a(x)).
\ee
since ${\widehat \ca}(x)$  is homeomorphic to the fiber $\pi^{-1}(x)$ of
$P^\infty$ we see then that there is a correspondence one to one between
IR's of $\ca_\infty$ and points of $P^\infty$. That
$\widehat{\ca}_\infty$
is homeomorphic to $P^\infty$ can be finally checked by observing that
closed sets of $P^\infty$ are also closed in the Hull-kernel topology
of $\ca_\infty$ and vice versa.

\bigskip

The generic case of a topological space $M$ can be treated by
reducing it to the one of the interval. Given a space $M$, consider a
sequence of finer and finer decompositions $\Sigma_n$ which satisfies
condition (\ref{reqseq}). The corresponding projective system of
noncommutative lattices $P^n(M)$ and its limit $P^\infty(M)$ were described
at the end
of section 2. In order to construct the dual inductive system,
consider again a cubic decomposition of $M$. The algebra $\ca_n(C_d)$,
of the noncommutative lattice $P^n(C_d)$ of a single cube given in eq.
(\ref{prod}), is simply the tensor product
\be
\ca_n(C_d) = \underbrace{\ca_n \otimes \ca_n \cdots \otimes
\ca_n}_d~,
\ee
where $\ca_n$ is the algebra of the noncommutative lattice $P^n$. This fact is
a
consequence of properties of IR's of the algebra of compact operators
and can be proven straightforwardly. The algebraic
operation of joining cubes is a direct
generalization of the one described in Example 3 of section 3.1.
We thus obtain a sequence of algebras $\{\ca_n(M)\}$ dual to
$\{P^n(M)\}$.
The inductive system is $\{\ca_n(M), \Phi_{(n,m)}(M)\}$ where, as
usual, the immersions $\Phi_{(n,m)}(M)$ are defined as the pull-backs
of the projections $\pi^{(m,n)}(M)$. Since the immersions
$\Phi_{(n,m)}(M)$ do not mix different factors in the direct sums of
algebras, the gluing procedure and the limiting procedure commute.
In order to find $\ca_\infty(M)$ and to show that is the dual of
$P^\infty(M)$ one can concentrate then on a single cube. For a single
cube, the following properties of $\ca_n(C_d)$ can be checked along
the same lines as for the interval,
\begin{itemize}
\item[1] The inductive limit $\bar{\ca}_\infty$  is the tensor product
       of the inductive limits $\ca_\infty$
\be
\bar{\ca}_\infty = \underbrace{\ca_\infty \otimes \ca_\infty
\cdots\otimes  \ca_\infty}_d
\ee
\item[2] The structure space of the algebra $\ca_\infty(x)$,
       generated by evaluating the elements of $\bar{\ca}_\infty$ at
       the point $x$ of the cube, is
       homeomorphic to the fiber of ${\bar P}^\infty$ at the point $x$.
\item[3] The center of $\bar{\ca}_\infty$ is the algebra of continuous
       functions of the cube.
\item[4] The section associated by the Dauns-Hofmann theorem to an element
of
       $\bar{\ca}_\infty$ is the operator valued function which
       defines the element itself.
\end{itemize}

\sxn {Direct Systems of Hilbert Spaces and their Limits}

In the previous sections we have considered the approximation of a topological
space $M$ and its algebra of continuous functions. In this section we want to
extend this analysis to the Hilbert space of square integrable functions
$L^2(M)$. We saw that the inductive limit of algebras, $\ca_\infty$,
consisted of
operator valued functions. We will see that the Hilbert space  $\ch^\infty$, on
which $\ca_\infty$ acts naturally, contains $L^2(M)$ as a subspace and is the
inductive limit of a sequence of Hilbert spaces $\ch^n$. On these spaces,
$\ch^n$, the algebras $\ca_n$ act naturally as bounded operators.

As before we start considering the interval $I$. The algebra
$\ca_\infty$ acts naturally on
the following Hilbert space $\ch^\infty$
\be
\ch^\infty = L^2(I) \bigoplus_{\alpha\in I_0}\ch_\alpha
\ee
where $\ch_\alpha$ are infinite dimensional Hilbert spaces. An element $\psi =
(f,\psi_\alpha)$ of $\ch^{\infty}$ consists then of a function $f$ of $L^2(I)$
and a countable collection of vectors $\psi_\alpha$ in $\ch_\alpha$ such that
\be
|| \psi || = \sum_\alpha <\psi_\alpha,\psi_\alpha > + ||f||^2
<\infty .
\ee
The space
$\ch^\infty$ is actually the inductive limit of the following sequence of
Hilbert spaces $\ch^n$ which is naturally associated to the noncommutative
lattices
$P^n$ and their algebras $\ca_n$. Each point $S^n_{m\alpha}$ of $P^n$
corresponds to an IR of $\ca_n$ on a Hilbert space $\ch^n_{m\alpha}$.
The space $\ch^n$
is then the direct sum of these $\ch^n_{m\alpha}$
\be
\ch_n = \bigoplus _{m\alpha}\ch^n_{m\alpha}.
\ee
where $\ch^n_{1\alpha} = {\bf C}$ and $\ch^n_{0\alpha}$ is infinite
dimensional.
The inner product of two elements $\psi^n =
\oplus_{m\alpha}\psi^n_{m\alpha}$ and $\phi^n =
\oplus_{m\alpha}\phi^n_{m\alpha}$ of $\ch_n$, is
\be
(\psi^n,\phi^n)_n = \sum_\alpha {1\over 2^{n-1}}
(\psi^n_{1\alpha})^* \phi^n_{1\alpha} +
\sum_\alpha (\psi^n_{0\alpha},\phi^n_{0\alpha})_{0\alpha}.
\ee
The algebra
$\ca_n$ can obviously be realized as bounded operators on $\ch_n$.
The isometric immersions $j^{n,n+1}$ of $\ch^n$ into $\ch^{n+1}$ defining
the inductive limit are given by the pull back
 of the projections $\pi^{(n+1,n)}$ from $P^{n+1}$ to $P^n$. Namely
\be
j^{n,n+1}(\oplus_{m\alpha} \psi^n_{m\alpha}) =
\oplus^{n+1}_{m\alpha} \psi^{n+1}_{m\alpha}
\ee
where
\be
\psi^{n+1}_{0\alpha} =
\left\{
\begin{array}{cc}
\psi^n_{0\beta}~~~ &if~~ \pi^{n+1,n}(S^{n+1}_{0\alpha}) = S^n_{0\beta}   \\
0  &if~~  \pi^{n+1,n}(S^{n+1}_{0\alpha}) = S^n_{1\beta}
\end{array}
\right.
\ee
\be
\psi^{n+1}_{1\alpha} = \psi^{n+1}_{1\alpha} ~~~~{\rm where} ~~
\pi^{n+1,n}(S^{n+1}_{1\alpha}) =  S^{n}_{1\beta}
\ee
The inductive limit of the inductive system of Hilbert spaces $\ch_n$
is obviously $\ch^\infty$. The space $L^2(I)$ is recovered as a subspace of
$\ch^\infty$.

The generalization to a generic space $M$ can be done following the argument
of the previous section. The construction is straightforward due to the fact
that tensor products and direct sums of algebras act respectively on tensor
products and direct sums of the corresponding Hilbert spaces.

\sxn{Conclusions}

In this paper we have discussed topological approximations of a
topological space $M$ by means of noncommutative lattices.
The space $M$ can be recovered from a projective limit $P^\infty(M)$
of a projective system associated with the noncommutative lattice
approximations.
In contrast to what happens for approximations with Hausdorff
lattices \cite{comp} the projective limit $P^\infty(M)$
is not universal and it is not `too different' from $M$ itself. The
latter can be naturally recovered from $P^\infty(M)$.

Any noncommutative lattice is the structure space of a noncommutative
\cstar of operator valued functions on it.
The collection of all such algebras is made into an inductive system
of $C^*$-algebras, from whose limit $\ca^\infty(M)$, the algebra
${\cal C}(M)$ of continuous functions on $M$ can be obtained as a
subalgebra. Again, in contrast to the Hausdorff approximations the
algebra $\ca^\infty(M)$ is not an universal object anymore.

\bigskip\bigskip

\noindent
{\large \bf Acknowledgements }

We would like to thank G. Marmo and P.
Michor for inviting us at ESI and all people at the Institute for the
warm and friendly atmosphere.
We thank A.P. Balachandran for many fruitful discussions and useful
advice. We also thank Mauro Carfora and Jerry Jungman for useful
discussions.

We thank the `Istituto Italiano di Studi Filosofici' in Napoli for
partial support.
The work of P.T-S. was also
supported by the Department of Energy, U.S.A. under
contract number DE-FG-02-84ER40173. The work of G.L. was
partially supported by the Italian `Ministero dell' Universit\`a e
della Ricerca Scientifica'.

\vfill\eject

\end{document}